\documentclass[twocolumn,showpacs,preprintnumbers,amsmath,amssymb,prb]{revtex4-1}
\usepackage{graphicx}
\usepackage{dcolumn}
\usepackage{bm}

\begin{document}
\title{Critical behavior of spin and chiral degrees of freedom in three-dimensional disordered
 XY models studied by the nonequilibrium aging method}
\author{F. Rom\'a}
\affiliation{Departamento de F\'{\i}sica, Universidad Nacional de San Luis. INFAP, CONICET. 
Chacabuco 917, D5700BWS San Luis, Argentina.}
\author{D. Dom\'{\i}nguez}
\affiliation{Centro At\'omico Bariloche and Instituto Balseiro, R8402AGP San Carlos de Bariloche, 
R\'{\i}o Negro, Argentina}

\date{\today}

\begin{abstract}
The critical behavior of the gauge-glass and the XY spin-glass models in three 
dimensions is studied by analyzing their nonequilibrium aging dynamics.  A new 
numerical method, which relies on the calculation of the two-time correlation and 
integrated response functions, is used to determine both the critical temperature 
and the nonequilibrium scaling exponents, both for spin and chiral degrees of freedom.
First, the  ferromagnetic XY model is studied to validate this nonequilibirum aging method 
(NAM), since for this nondisordered system we can compare with known results obtained 
with standard equilibrium and nonequilibrium techniques. When applied to the case of 
the gauge-glass model, we show that the NAM allows us to obtain precise and reliable 
values of its critical quantities, improving previous estimates.  The XY spin-glass model 
with both Gaussian and bimodal bond distributions, is analyzed in more detail. The spin and 
the chiral two-time correlation and integrated response functions are calculated in our 
simulations.  The results obtained mainly for Gaussian and,
to a lesser extent, for bimodal interactions, support the existence of a spin-chiral 
decoupling scenario, where the chiral order occurs at a finite temperature while the spin 
degrees of freedom order at very low or zero temperature.  

\end{abstract}


\maketitle

\section{\label{S1}INTRODUCTION}

Critical phenomena in disordered and frustrated systems such as spin glasses 
are one of the hardest problems in the statistical mechanics theory.  Extensive 
numerical simulations of the three-dimensional (3D) Ising spin-glass model \cite{Ballesteros2000,Katzgraber2006} 
for example, have confirmed that this system undergoes a transition at a 
finite critical temperature, but the nature of this phase transition is still 
under discussion.  More controversial is the case of those models with 
XY-type spins.  While early studies of 3D XY spin glasses indicated the existence 
of a transition at very low or zero temperature,\cite{Morris1986,Jain1986} 
subsequent numerical simulations reveal other possible scenarios. 

In 3D XY spin-glass models, two kinds of symmetries can be broken because 
the Hamiltonian is invariant under both, the usual continuous ``spin'' O(2) 
symmetry associated to a global rotation, and the ``chiral'' symmetry 
associated to the reflection of XY spins about an arbitrary direction.  Following 
the seminal work of Villain,\cite{Villain1977} some works have shown the possibility 
that the spin order and the chiral order occur at different critical temperatures, 
which is known as the ``spin-chiral decoupling'' scenario.\cite{Kawamura1987,Kawamura2001,Obuchi2013}  
Nevertheless, other numerical simulation studies suggest the existence of 
a single finite critical temperature at which both symmetries  are simultaneously 
broken.\cite{Maucourt1998,Granato2001,Granato2004a,Granato2004b,Lee2003,Yamamoto2004,Chen2009} 
Recently, an even  more complex scenario  
has been proposed, in which the lower critical dimension 
of the model could be close or equal to three, and then the finite-temperature 
phase transition should be removed by fluctuations. \cite{Pixley2008}  In this 
case only a marginal behavior for low temperature and large sizes should be expected.  
For a recent review interested readers can refer to Ref. \onlinecite{Kawamura2010}. 

In all these studies, extensive equilibrium and nonequilibrium simulations were 
performed to solve this intricate subject. The lack of a consensus is probably due 
to the hardness of the problem.  Recently, a new nonequilibrium aging method
(NAM) was proposed to analyze such situations. \cite{Roma2010}  Calculating the 
correlation and the integrated response functions for different temperatures 
and waiting times, it is possible to determine both the critical temperature and 
the nonequilibrium scaling exponents.  Using this very sensitive technique, it has 
been possible to corroborate that the universality class of the equilibrium phase 
transition in 3D Ising spin glasses does not depend on the exact form of the bond 
distribution.\cite{note1}  

With the aid of the nonequilibrium aging method proposed in Ref. \onlinecite{Roma2010}, 
in this work we study the equilibrium critical behavior of three different 3D 
XY systems: the classical ferromagnetic XY model, the gauge-glass model \cite{Huse1990} 
and, in more detail, the XY spin-glass model with both Gaussian and bimodal bond 
distributions.  The first two systems of this class are mainly analyzed to show 
that the NAM numerical technique is useful to locate the critical point of either an 
ordered or a disordered model with XY-type spins.  Nevertheless, for the gauge-glass 
model, we report precise values of both the critical temperature and the nonequilibrium 
exponents, improving previous estimates.  
On the other hand, for 
the XY spin-glass models, our results strongly support the existence of a spin-
chiral decoupling scenario where the chiral order occurs at finite temperature 
while the spins order at very low or zero temperature.   

The paper is organized as follows. In Sec. \ref{method}, we present  
the nonequilibrium aging method and the simulation scheme. Then, in Sec. \ref{XYF-GG} 
we analyze the classical ferromagnetic XY and the gauge-glass models.  Section 
\ref{XYglass} is devoted to the study of the XY spin-glass system with 
both Gaussian and bimodal distributions of bonds.  Finally, in Sec. \ref{DC} 
we discuss our results.

\section{\label{method}  The NONEQUILIBRIUM AGING METHOD and simulation scheme}

Numerical nonequilibrium relaxation methods are frequently used to analyze equilibrium phase 
transitions.\cite{Ozeki2007} The simplest protocol consists in preparing the system 
at time $t=0$, in a fully-ordered state and then the dynamics is simulated with a 
standard Monte Carlo algorithm.  The critical temperature $T_c$ is estimated as the 
temperature at which, in the asymptotic regime, the order parameter follows a power-law
dependence in time. 
This method is appropriate to study a wide variety of systems,
since 
the slow dynamics present in disordered and frustrated systems favors the application 
of such nonequilibrium technique.
However, due mainly to the fact that different 
observables seem to decay by a power law in a relatively wide interval of temperatures, 
it is not possible to determine an accurate value of $T_c$ with these methods.  
Additional scaling analysis 
of the order parameter or susceptibility has been used to improve the resolution of 
these methods.\cite{Bernardi1996,Ozeki2001,Nakamura2003} In addition, recently it 
has been proposed a new technique based on the divergence of the relaxation time 
approaching the critical point. \cite{Lippiello2010}  

The nonequilibrium aging method (NAM) proposed in Ref.\onlinecite{Roma2010} is a general technique 
that allows us to overcome these difficulties. First, a typical protocol is used which 
consists on a quench at $t=0$ from a disordered state ($T \to \infty$) to a low 
temperature $T$. From this initial condition, the system is simulated by a Model A 
dynamics,\cite{Hohenberg1977,Folk2006} in this case a standard Glauber dynamics.  Then, a given 
autocorrelation function $\mathcal{C}(t,t_w)$ and its associated integrated autoresponse 
function $\varrho(t,t_w)$ (or merely the correlation and the integrated response functions), 
which depend on both, the waiting time $t_w$ when the measurement begins and a given time 
$t>t_w$, are calculated for different temperatures and different values of $t_w$.  
Later, for each set of curves of a given $T$, a data collapse analysis is performed 
based on the following scaling relations for the correlation function
\begin{equation}
\mathcal{C}(t,t_w) = {t_w}^{-b} f_\mathcal{C} (t/t_w),  \label{scalingCorr}
\end{equation}
and for the integrated response function 
\begin{equation}
\varrho(t,t_w) = {t_w}^{-a} f_{\varrho} (t/t_w) , \label{scalingResp}
\end{equation}
 where $b$ and $a$ are two nonequilibrium exponents, and $f_{\mathcal{C}/\varrho}$ 
are two different scaling functions.\cite{Pleimling2005,Godreche2002}  Relations (\ref{scalingCorr}) 
and (\ref{scalingResp}) are expected to be correct only at the critical point with 
$b=a$ and for $t_w \ll t-t_w $. \cite{Janssen1989,Calabrese2005}  The best data 
collapse, and therefore the best candidate values for  $b$ and $a$ at a 
given $T$,  is obtained by minimizing 
the sum of squared differences between all pairs of curves within a given range 
of times (see details in the following).  Then, to identify $T_c$, simply we chose the 
temperature for which the condition $b=a$ is fulfilled.  Note that, for temperatures 
above or below $T_c$, the method provides only pseudo-exponents because other scaling 
relations, or a separation of the correlation and the response functions in their 
corresponding stationary and aging terms,\cite{Lippiello2006} are required.  
Throughout this work, we will use the NAM to determine the critical temperature 
and the nonequilibrium exponents for the different XY models.   

For systems of $N$ XY-type spins, the spin correlation function is defined as 
\begin{eqnarray}
C(t,t_w) &=& \frac{1}{N} {\left[ \left \langle  \sum_{i=1}^N \mathbf{S}_i(t) \cdot \mathbf{S}_i(t_w)  \right\rangle_0  \right]} \nonumber \\
&=&\frac{1}{N} {\left[ \left \langle \sum_{i=1}^N \cos \left( \theta_{i}(t)-\theta_{i}(t_w) \right) \right \rangle_0\right]}, \label{corrS}
\end{eqnarray}
where the sum runs over the sites of the lattice, $\mathbf{S}_i =(\cos\theta_i,\sin\theta_i)$ 
are classical two-dimensional spins of unit length, $\theta_{i}$ are angular variables, 
$\langle ... \rangle_0$ indicates an average over different thermal histories 
(different initial configurations and realizations of the thermal noise) and $[...]$ 
represents an average over different disordered samples.   On the other hand, 
adding to the Hamiltonian of the system, $\mathcal{H}$, a perturbation of the form
\begin{equation}
\mathcal{H}_p = - \sum_{i=1}^N \mathbf{h}_i \cdot \mathbf{S}_i , \label{perS}
\end{equation} 
where $\mathbf{h}_i$ is a local external field, and switching on this perturbation 
only for times $t < t_w$, it is possible to calculate a (reduced) spin integrated 
(thermoremanent) response function as \cite{Godreche2002}    
\begin{equation}
\rho(t,t_w) = \frac{T}{N} \left[ \sum_{i=1}^N  \frac{ \partial \langle \cos \theta_i (t) \rangle_h}{\partial h_{x,i}}\Bigg|_{h=0} + \frac{\partial \langle \sin \theta_i (t) \rangle_h}{\partial h_{y,i}}\Bigg|_{h=0} \right],  \label{respS}
\end{equation}
where now $\langle ... \rangle_h$ indicates an average over thermal histories 
of the perturbed system.  

At thermodynamic equilibrium, both the correlation (\ref{corrS}) and the 
integrated response (\ref{perS}), depend on $\tau=t-t_w$ and are 
related through the fluctuation-dissipation theorem (FDT)    
\begin{equation}
\rho(t-t_w) = C(t-t_w).  \label{FDT_S}
\end{equation}
For a nonequilibrium process, however, the FDT is not fulfilled.  Nevertheless, it has 
been proposed that a generalized quasi-fluctuation-dissipation theorem (QFDT) \cite{Cugliandolo1994} 
of the form 
\begin{equation}
\frac{\partial \rho(t,t_w)}{\partial t_w} = X(t,t_w) \frac{\partial C(t,t_w)}{\partial t_w}, \label{QFDT_S}
\end{equation} 
where $X(t,t_w)$ is the fluctuation-dissipation ratio, must be obeyed by any physical model.  
For short times ($\tau \ll t_w$), 
a system is in the quasi-equilibrium regime ($X \approx 1$) and therefore a 
parametric plot of $\rho$ vs. $C$ should show a straight line of slope $1$. 
On the other hand, for long times ($t_w \ll \tau$) we can write \cite{Godreche2002}  
\begin{equation}
\rho(t,t_w) = X_\infty C(t,t_w), \label{QFDT_S2}
\end{equation} 
where
\begin{equation}
X_\infty = \lim_{t_w\rightarrow\infty} \lim_{t\rightarrow\infty} X(t,t_w).  \label{XSinfty} 
\end{equation}
In a critical quench, it is expected that the spin correlation (\ref{corrS}) 
and the spin integrated response (\ref{respS}) follow, respectively, 
the scaling relations (\ref{scalingCorr}) and (\ref{scalingResp}).       

For the XY spin-glass systems,  we can also define the local chirality of each 
square plaquette $\alpha$ by \cite{Kawamura2001}  
\begin{equation}
\kappa_\alpha = \frac{1}{2 \sqrt{2}} \sum_{(i,j) \in \alpha} \mathrm{sgn} (J_{ij}) \sin(\theta_i - \theta_j), 
\end{equation}
where the sum runs over the four bonds of strength $J_{ij}$ surrounding the 
plaquette $\alpha$ in a clockwise direction.  Then, the chiral correlation 
can be written as     
\begin{equation}
C_\kappa(t,t_w) = \frac{1}{3N} {\left[ \left \langle  \sum_\alpha \kappa_\alpha(t) \kappa_\alpha(t_w) \right\rangle_0  \right]} . \label{corrC}
\end{equation}
Here the sum is taken over all the $3N$ plaquettes. To calculate a response 
function we need to add to the Hamiltonian a different perturbation   
\begin{equation}
\mathcal{H}_{p \kappa} = - \sum_\alpha f_\alpha \kappa_\alpha, \label{perC}
\end{equation}
where $f_\alpha$ is a fictitious external field which couples to the chirality 
$\kappa_\alpha$. Then, the corresponding (reduced) chiral integrated (thermoremanent) response 
function is given by 
\begin{equation}
\rho_\kappa(t,t_w) = \frac{T}{3N} \left[ \sum_\alpha  \frac{ \partial \langle \kappa_\alpha \rangle_f}{\partial f_\alpha}\Bigg|_{f=0} \right].  \label{respC}
\end{equation}
In this equation, $\langle ... \rangle_f$ indicates an average over thermal 
histories when the system is perturbed by (\ref{perC}). At  equilibrium 
the FDT relation is satisfied     
\begin{equation}
\rho_\kappa(t-t_w) = C_\kappa(t-t_w),  \label{FDT_C}
\end{equation}
but far away from this condition, it is expected that in the quasi-equilibrium regime 
the FDT still works while for long times ($t_w \ll \tau$) a QFDT holds,  
\begin{equation}
\rho_\kappa(t,t_w) = X_{_\kappa \infty} C_\kappa(t,t_w). \label{QFDT_C2}
\end{equation} 
As before, in a critical quench the chiral functions (\ref{corrC}) and (\ref{respC}) should follow, 
respectively, the scaling relations (\ref{scalingCorr}) and (\ref{scalingResp}).  To avoid confusion, 
we will denominate the corresponding nonequilibrium exponents as $b_\kappa$ 
and $a_\kappa$, reserving $b$ and $a$ for the spin function (\ref{corrS}) and (\ref{respS}).     

In this work, instead of performing additional simulations with applied fields 
of small strength, the integrated response functions (\ref{respS}) and 
(\ref{respC}) were calculated for infinitesimal perturbations using the 
algorithm proposed in Refs.~\onlinecite{Chatelain2003,Ricci-Tersenghi2003}.  
This technique permits us to determine the two correlation functions and the 
two response functions in a single simulation of the unperturbed system.   
Thus, we can obtain  reliable values of both exponents $b$ and $a$ (or $b_\kappa$ and $a_\kappa$), 
making possible the realization of this nonequilibrium aging method.  
Further details of  this algorithm are given in the Appendix.   

For the Monte Carlo simulation we have used a standard scheme, where local 
changes in the phases $\theta_{i} \to \theta'_{i}$ are accepted with a
probability given by the Glauber rate
\begin{equation}
\omega( \theta_{i} \to \theta'_{i} ) = \frac{\exp \left( - \beta \Delta \mathcal{H} \right) }{1 + \exp
\left( - \beta \Delta \mathcal{H} \right)}. \label{rate}
\end{equation}
Here $\beta$ is the inverse temperature and $\Delta \mathcal{H}$ is the 
energy difference corresponding to the proposed phase change. 
Typically, in equilibrium 
simulations, the acceptance window for $\theta'_{i}$ is chosen less 
than $2\pi$ and dependent on temperature in order to optimize the updating 
procedure. Since we are interested in studying a nonequilibrium process, 
for local phase changes we will use the full $2\pi$ acceptance angle window 
for the new phases $\theta'_{i}$, 
following the criterion used in Ref.~\onlinecite{Katzgraber2004,Katzgraber2005}. 
This is done in order to avoid the possibility that a limited acceptance 
angle might introduce an artificial temperature dependence in the relaxation. 
In all cases, cubic lattices of linear size $L=50$ with full periodic boundary conditions 
were simulated and the disorder average was performed over $10^4$ different samples 
for each temperature (for the ferromagnetic XY model, a nondisordered system, we have 
carried out $10^4$ independent thermal histories).

\section{\label{XYF-GG} XY models with spin symmetry breaking}

In this section we study the critical behavior of the classical ferromagnetic 
XY and the gauge-glass models.  These systems are known to have a single phase
transition where the spin U(1) symmetry is broken. We first study
the well-known 3D ferromagnetic XY model  in order to validate 
the NAM, showing that this technique works well in a 
nondisordered system, allowing us to obtain a good estimate of the
dynamical exponent $z$. Afterwards, we  study the disordered gauge-glass model, 
showing that the NAM can
give a precise estimate of its critical temperature and dynamical coefficients.      

\subsection{\label{XYF} Ferromagnetic XY model}

In order to validate the NAM, we study the critical 
behavior of the 3D ferromagnetic XY model. The Hamiltonian 
of this system is given by
\begin{equation}
\mathcal{H}_{XY} = - \sum_{( i,j )} \mathbf{S}_i \cdot \mathbf{S}_j , \label{HamXYferro}
\end{equation}
where the sum runs over all nearest-neighbor pairs $(i,j)$ on a cubic lattice 
of $N=L^3$ spins.  In terms of the angular variables $\theta_{i}$, the 
Hamiltonian can be written as  
\begin{equation}
\mathcal{H}_{XY} = - \sum_{( i,j )} \cos(\theta_{i} - \theta_{j}). \label{HamXYferro2}
\end{equation}
The critical behavior of this model is well known.  From high-precision 
Monte Carlo simulations it has been possible to estimate the critical temperature 
$T_c = 2.20167(10)$ \cite{Gottlob1993} and the critical exponent $\eta = 0.0381(2)$.\cite{Hasenbusch1999} 
There are also some less accurate
estimates of the critical dynamic exponent as being $z_c \approx 2$,\cite{Minnhagen2001}
obtained using relaxation dynamics with periodic boundary conditions 
(the result was found to depend on the  boundary condition in the work cited above). 

\begin{figure}[t!]
\includegraphics[width=6.5cm,clip=true]{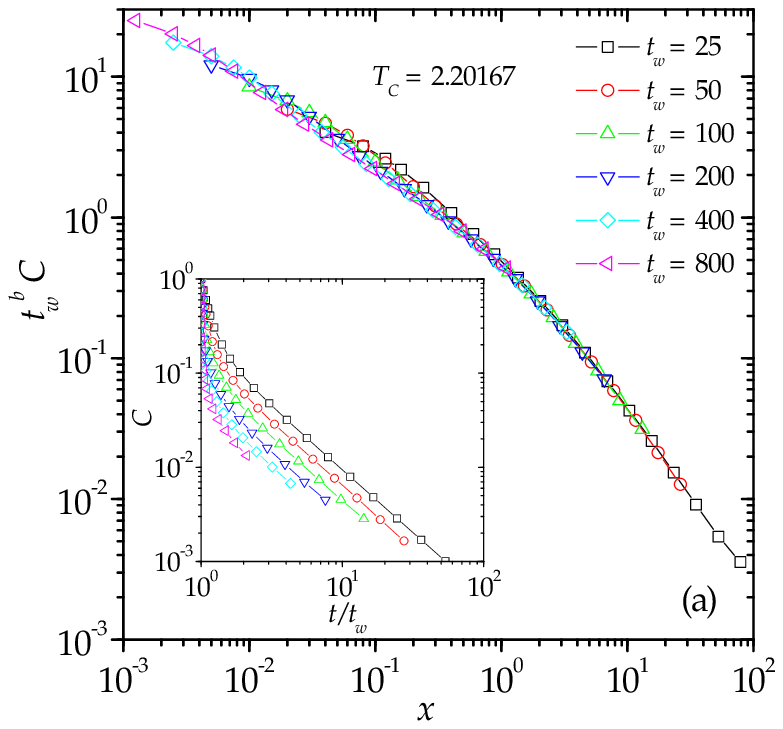}
\includegraphics[width=6.5cm,clip=true]{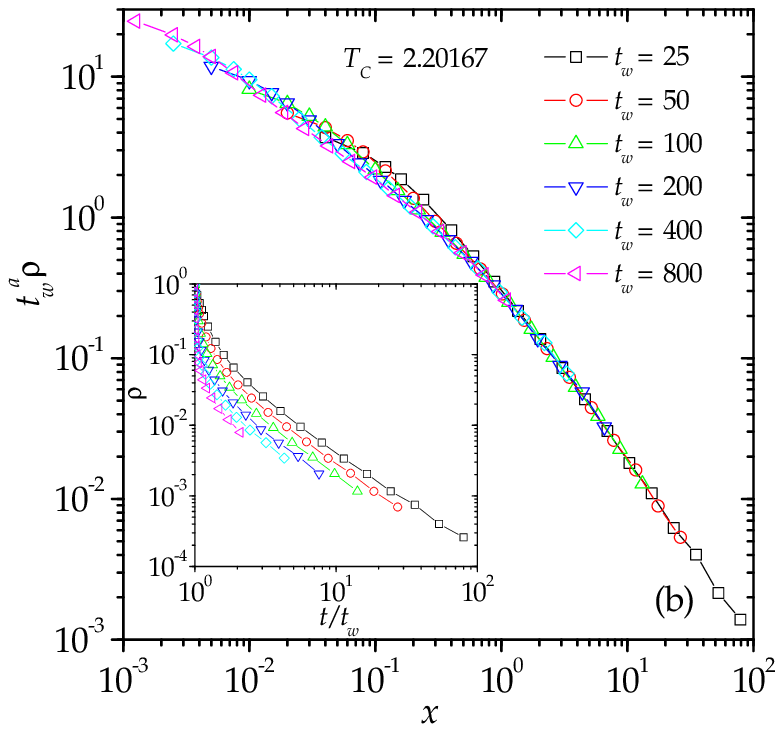}
\includegraphics[width=6.5cm,clip=true]{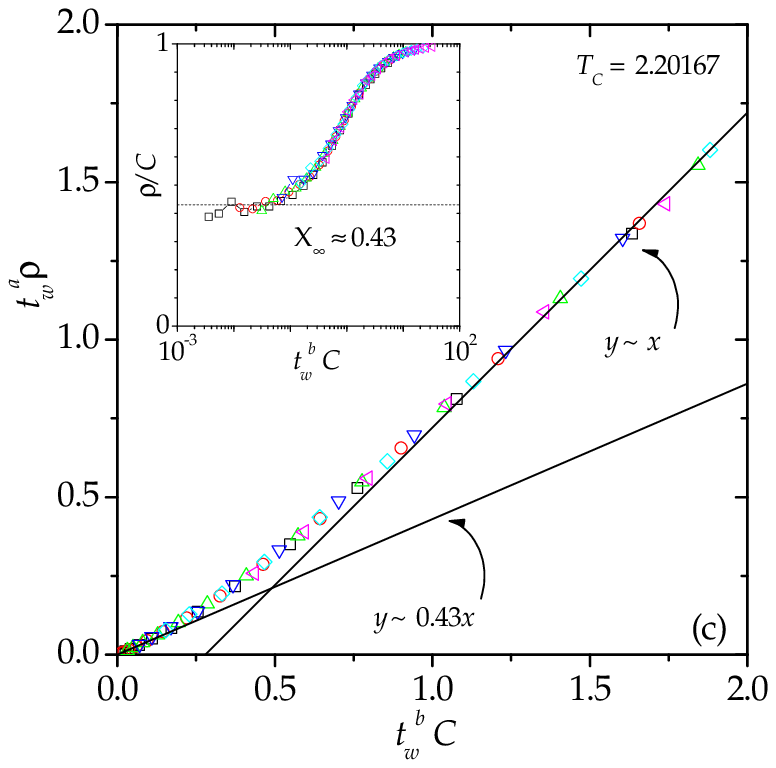}
\caption{\label{figure1} (Color online) Ferromagnetic XY model. 
Panels (a) and (b) show, respectively, the data collapse with $b=a=0.521$ 
of the spin correlation and the spin integrated response at $T=2.20167$ 
for different $t_w$ as indicated. Insets show the corresponding 
curves as function of $t/t_w$.  (c) FDT plot.  Inset shows the plot of 
$\rho/C$ vs. $t_w^b C$ and the value of $X_\infty \approx 0.43$.}
\end{figure}

\begin{figure}[t!]
\includegraphics[width=6.5cm,clip=true]{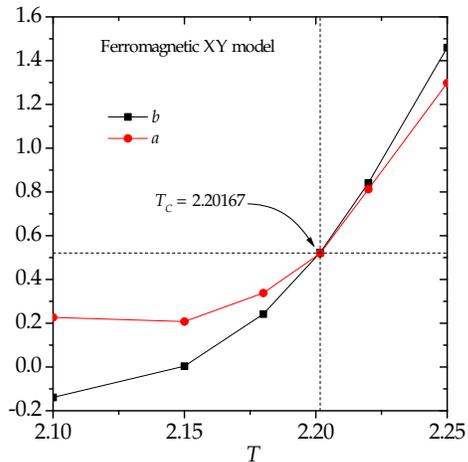}
\caption{\label{figure2} (Color online) The values of the exponents $b$ and $a$ vs. $T$,
obtained with the NAM for the ferromagnetic XY model.}
\end{figure}

We start verifying that the behavior 
described by Eqs.(\ref{scalingCorr}) and (\ref{scalingResp}) for a critical point, is followed
by this model  at the known critical temperature $T_c= 2.20167(10)$.\cite{Gottlob1993}
Figures \ref{figure1} (a) and (b) show, respectively, the data collapse of 
the spin correlation (taken as $t_w^bC$) and the spin integrated response functions
(taken as $t_w^a\rho$), for waiting 
times $t_w=25, 50, 100, 200, 400$ and $800$ at  $T = 2.20167 \approx T_c$, 
plotted as function of the variable $x=\tau/t_w=t/t_w-1$. 
For each set, to obtain a good data collapse it is necessary to minimize 
the sum of squared differences between all pairs of curves within a  
range $[x_0,x_m]$, with $x_m$  the maximum value of $x$ calculated.  
The starting $x_0$ is chosen as large as possible so that the scaling region ($x \gg 1 $) 
can be reached, but at the same
time not so large as to have an interval of length $x_m-x_0$ with enough 
data points to have a significant statistics. Furthermore, data 
collapses with a varying number $n$ of curves with different waiting
times $t_w$ are attempted. The number $n$ means to
have considered in the data collapse the $n$ largest values of $t_w$ available 
(ranging from $n=6$ for $t_w=25, 50, 100, 200, 400, 800$
to $n=2$ for $t_w= 400, 800$). Small $n$ gives more weight to
the large $t_w$ data but at the expense of poorer statistics for
the data collapse. Varying $x_0$ and $n$ we obtain several pairs of exponents $b$ and $a$.  
We then compare the $b$ and $a$ obtained using different $x_0$ and $n$,
taking the ones with the smallest $x_0$ and the the largest $n$, such that 
when increasing $x_0$ or decreasing $n$ from this point,
the exponent values do not change appreciably.

To obtain the data collapses shown in Figs. \ref{figure1} (a) and (b) it was enough 
to choose $x_0 = 0.5$ and to take the $t_w$-curves with $t_w=100 - 800$ (i.e., $n=4$). 
Then, with the minimum squares method  the exponents values $b=0.5225$ and 
$a=0.5195$ were obtained. At the critical temperature $T_c$, we should have $a=b$, and indeed the
values of $a$ and $b$ obtained are very close.
Since we are considering the data at a temperature equal to the best known estimate of $T_c$,
we  use a common mean value of $c=(b+a)/2=0.521$ to collapse simultaneously the
sets of curves shown in Figs. \ref{figure1} (a) and (b).  
In Fig. \ref{figure1} (c) we show the corresponding ``FDT plot'', where
$t_w^a \rho$ vs. $t_w^b C$ are plotted.  As mentioned in 
the previous section, for times such that $\tau\ll t_w$, the system is in the quasi-equilibrium 
regime and $X \approx 1$.  This behavior is observed in Fig. \ref{figure1} (c) 
where for large values of $C$ and $\rho$, the slope in the FDT plot tends to $1$.  
On the other hand, for times $\tau\gg t_w$ the QFDT holds and the slope near to the origin is equal 
to $X_\infty$.  From (\ref{QFDT_S2}) and (\ref{XSinfty}) we see that
\begin{equation}
X_\infty = \lim_{t_w\rightarrow\infty} \lim_{t\rightarrow\infty} \frac{\rho}{C}.  \label{XSinfty2} 
\end{equation} 
The inset in Fig. \ref{figure1}(c) shows that this ratio tends to $X_\infty \approx 0.43$.
This same value was obtained in Ref. \onlinecite{Abriet2004} where the
nonequilibrium dynamics of the XY model was extensively studied.            

Now, we follow in its full extent 
the procedure of the NAM described
in  Sec. \ref{method}, analyzing the behavior of correlations
and response functions in a range of temperatures. We simulate 
the system for six temperatures 
between $T=2.1$ and $2.25$, and waiting times $t_w=25, 50, 100, 200, 400$ and $800$.   
Figure \ref{figure2} shows the exponents $b$ and $a$ vs. $T$, obtained fitting 
Eqs.(\ref{scalingCorr}) and (\ref{scalingResp}) 
using the same parameters as above: $x_0 = 0.5$ and curves with $t_w=100 - 800$ ($n=4$). 
We see that the $a(T)$ data and the $b(T)$ data cross at 
approximately the above used $T_c$, where the condition $b=a$ should be fulfilled.
This confirms that the NAM introduced in Sec. \ref{method}
is reliable.  Moreover, if the data collapse were performed for 
larger ranges of $x$, for example starting from $x_0=0$, a good value of the
critical temperature but bad values for the exponents $b$ and $a$ are obtained.       

Alternatively, one can quantify the goodness of the power-law collapses
by calculating the sum of squared differences between all pairs of curves,
$\Delta^2$. \cite{Roma2010}  A plot of this quantity as a function of $T$ has a
a well-defined minimum at $T_c$, since, as showed above, the correlation and 
the integrated response functions display a power-law behavior in a very narrow 
range of $T$ around $T_c$. A very similar behavior was shown in the Fig. 1(d) in Ref. \onlinecite{Roma2010}
for the two-dimensional ferromagnetic Ising model.  In disordered systems 
(as the cases that will be discussed in the following sections) 
the dynamical correlation and response functions can be fitted with reasonably
good power-law decays within a relatively wide interval of temperatures
and studying quantities such as $\Delta^2$ is not useful in practice.
On the other hand, the coincidence of $b=a$ is a stronger requirement than
a minimum in $\Delta^2$, giving a better evidence for the existence
of a critical point, which we find more useful in disordered systems. 

Next, to calculate suitable error bars for the critical 
temperature and the common exponent $c$, which is a reasonable estimate of the 
nonequilibrium critical exponent $b$ (or $a$), we consider the dispersion 
of the coordinate values of the crossing point 
calculated for different ``acceptable'' data collapses 
(those for which the parameter $x_0$ is greater than but close to $0.5$).  
In this way we obtain the estimates $T_c=2.20(1)$ and $b=0.52(1)$.  Note that this critical 
temperature is compatible with the more precise value $T_c = 2.20167(10)$.\cite{Gottlob1993}           
The exponent $b$ can be related with the critical dynamical exponent $z_c$ 
from the relation \cite{Godreche2002}
\begin{equation}
b=\frac{(d-2+\eta)}{z_c}. \label{Eq-b}
\end{equation}
Using the known result $\eta = 0.0381(2)$, and the calculated $b$, we
obtain $z_c=2.00(2)$, which is in good agreement with
the numerical results of Ref. \onlinecite{Minnhagen2001} and with
epsilon expansion calculations of $z_c$ for model A dynamics,\cite{Hohenberg1977,Folk2006}
that give $z_c=2.022$ for this model.

Our analysis of the ferromagnetic XY model corroborates that the NAM works well. 
However, as a procedure to obtain the critical temperature
$T_c$, this method is not competitive
when compared to the standard equilibrium and nonequilibrium techniques \cite{Ozeki2007} for
nondisordered systems, since the error bars in $T_c$ calculated here are comparatively very large.
Nevertheless, as we shall see in the next subsection, the NAM 
will allow us to calculate reliable values of the critical quantities for disordered systems.

\subsection{\label{GG} Gauge-glass model}

The 3D gauge-glass (GG) model \cite{Huse1990,Reger1991,Moore} is  a paradigmatic model for the 
vortex glass phase transition in superconductors.\cite{review1,review2}  Its Hamiltonian is given by
\begin{equation}
\mathcal{H}_{\rm GG} = - \sum_{( i,j )} \cos(\theta_{i} - \theta_{j} - A_{ij}),
\label{HGG}
\end{equation}
expression that is similar to the ferromagnetic Hamiltonian (\ref{HamXYferro}), but with additional
quenched random variables $A_{i,j}$ which are drawn from a uniform distribution in the $[0,2\pi]$ 
interval under the constraint $A_{ij}=-A_{ji}$.  We simulate the GG system for seven temperatures between 
$T=0.42$ and $0.48$, and waiting times $t_w= 50, 100, 200, 400$, $800$, $1600$, $3200$ and $6400$.

\begin{figure} [t!]
\includegraphics[width=6.5cm,clip=true]{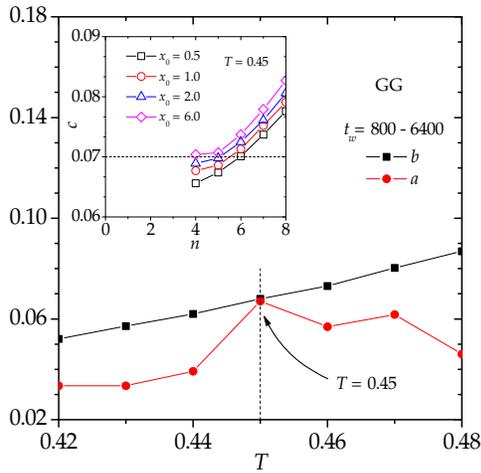}
\caption{\label{figure3} (Color online) The values of the exponents $b$ and $a$ vs. $T$,
obtained for the GG model.  The parameters used were $x_0=1$ and the range $t_w = 800 - 6400$ ($n=4$) 
as indicated. The inset shows the dependence of the exponent $c$ with the quantity $n$, 
for different values of $x_0$ at $T=0.45$.}
\end{figure}

Figure \ref{figure3} shows the curves of $b$ and $a$ vs. $T$, which were obtained by choosing 
$x_0 = 1$ and the range $t_w = 800 - 6400$ ($n=4$). We see that  
that the condition $b=a$ is fulfilled as a contact point of the $a(T)$ curve with
the $b(T)$ curve at $T \approx 0.45$. This is different
from the previous case where the condition $b(T_c)=a(T_c)$ was found as 
a crossing point of the two curves. 
This seems to be a common feature of disordered systems:
a similar behavior was observed in the Ising spin glasses, where 
the exponent curves do not cross but coincide  at a contact point  at $T_c$.\cite{Roma2010}  
Analyzing  different values of $x_0$ between $0.5$ and $6$, and different ranges of $t_w$ 
up to $t_w = 800 - 6400$ (i.e. from $n=8$ to $4$), 
we obtain a very stable contact point close to $0.45$ and
for this reason we regard this temperature as the critical one, $T_c = 0.450(3)$. 
As before, to calculate a suitable error bar we have considered the dispersion 
of the contact point for different ``acceptable'' data collapses. 
    
\begin{figure} [t!]
\includegraphics[width=6.5cm,clip=true]{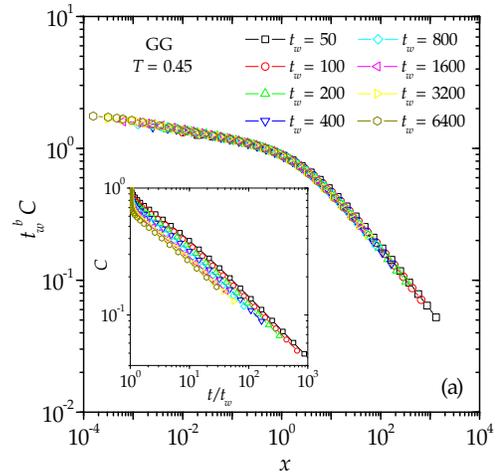}
\includegraphics[width=6.5cm,clip=true]{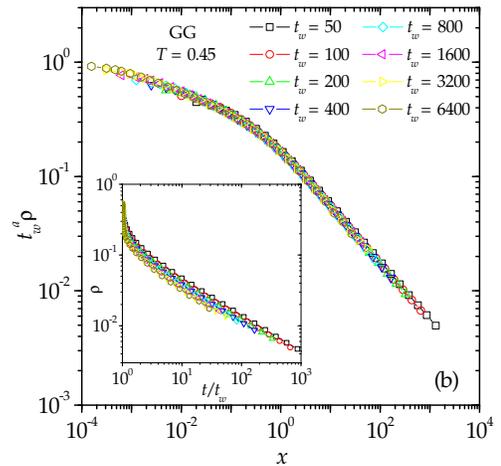}
\includegraphics[width=6.5cm,clip=true]{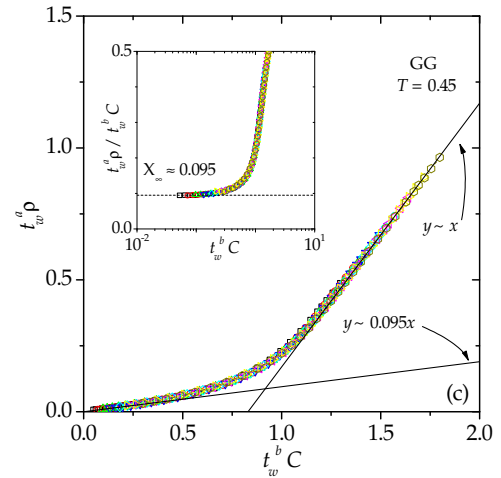}
\caption{\label{figure4} (Color online) GG model. Panels (a) and (b) show, respectively, 
the data collapse with $b=a=0.07$ of the spin correlation and the spin integrated 
response at $T=0.45$ for different $t_w$ as indicated. Both insets show the corresponding 
curves as function of $t/t_w$.  (c) FDT plot.  Inset shows the plot of 
$\rho/C$ vs. $t_w^b C$ and the estimated value of $X_\infty \approx 0.095$.}
\end{figure}

Unlike the  behavior observed for the ferromagnetic model,
we find that the contact point tends to move towards lower 
values of $b$ and $a$, when we 
plot curves $a(T),b(T)$ obtained from fits with increasing
$x_0$ or by removing progressively the curves
with smaller $t_w$ (decreasing $n$).
This is shown in the inset of Fig. \ref{figure3}
where we plot $c=(a+b)/2$ at $T=0.45$  for different values of
 $x_0$ and $n$.
Then, even when we find clear evidence that there is a critical point
(the condition $b=a$ is fulfilled in each of these cases), the value of the
exponent $b=a$ is not fully converged for the time scales considered. 
To obtain a value for this exponent, we consider that the best estimate corresponds to
the fits that are more weighted at the largest time scales $t$ and $t_w$;
i.e., the limit value of $c$ when increasing $x_0$ and decreasing $n$. 
With this methodology we estimate $c = 0.070(4)$ from the data
plotted in the inset of Fig. \ref{figure3}.   

Figures \ref{figure4} (a) and (b) show the curves of the spin correlation and
the spin integrated response, where a value of $b=a=0.07$ was used 
to collapse simultaneously both sets of curves.  In addition, 
Fig. \ref{figure4}(c) shows the corresponding FDT plot.  As before, we observe that 
for short times the system is in the quasi-equilibrium regime and $X \approx 1$, 
but for long times the FDT is violated and the slope near to the origin is equal 
to $X_\infty \approx 0.095$. 

The 3D GG model has been widely studied by numerical techniques. 
In an early equilibrium simulation study, Reger {\em et al.} \cite{Reger1991} found 
$T_c = 0.45(5)$ and later, using the most efficient exchange Monte Carlo algorithm, Olson and 
Young \cite{Olson2000} estimated $T_c = 0.47(3)$.  More recently, using a variety of numerical methods 
Katzgraber and Campbell \cite{Katzgraber2004} obtained $T_c = 0.46(1)$, the same value reported by 
Alba and Vicari \cite{Alba2011} from Monte Carlo simulations and finite-size scaling analyses. 
Note that our estimation of the critical temperature, $T_c = 0.450(3)$, is 
compatible with these values and is a bit more precise.   

On the other hand, our estimation of the nonequilibrium exponent $c = 0.070(4)$,
is close to but slightly larger than the value of $b$ reported by us in a previous work,\cite{Roma2008}
$b = 0.06(1)$. However, in Ref. \onlinecite{Roma2008} the $b$ exponent was calculated differently 
by fitting the correlation function for short times with a power law, a method very imprecise. 
Finally, in this work we have determined that $X_\infty \approx 0.095$, 
but in Ref. \onlinecite{Roma2008} was reported a larger value of $X_\infty \approx 0.12$.
This difference is clearly attributed to the improved statistics in this
work: in the cited reference averages were carried out over only $60$ samples, 
while here $10^4$ different realizations of disorder were used for each temperature.    

\section{\label{XYglass} XY spin-glass models}

The 3D XY spin-glass model is a paradigmatic model for disorder and frustrated magnetic materials 
with an easy-plane-type anisotropy.\cite{Murayama1986,Katsumata1988,Ito1992,Kawano1993,Mathieu2005,Kawamura2011,Yamaguchi2012}  
Another experimental realization of this system are granular cuprate superconductors consisting 
of a network of sub-micron-size superconducting grains randomly connected through $\pi$-Josephson 
junctions.\cite{Dominguez1993,Matsuura1995,Kawamura1995,Kawamura1996-97,Gardchareon2003}

The Hamiltonian of the 3D XY spin-glass model is 
\begin{equation}
\mathcal{H}_{XYglass} = - \sum_{( i,j )} J_{ij} \cos(\theta_{i} - \theta_{j}),
\label{XYglassH}
\end{equation}
where as before the sum runs over all nearest-neighbor pairs $(i,j)$ on a cubic lattice.  The quenched 
random bonds $J_{i,j}$ are drawn from a distribution with mean zero and variance one. 
We present a detailed analysis  with both,
Gaussian and bimodal $\pm J$ bond distributions.  
For clarity, we name XYG and XYB  the XY spin-glass models with
a Gaussian and a bimodal $\pm J$ (with $J=1$) bond distributions, respectively. 
Spin and chiral observables (correlation and 
integrated response functions) are calculated for several temperatures in a wide range, and
for waiting times $t_w= 50, 100, 200, 400$, $800$, and $1600$.
  
As mentioned in the Introduction, different scenarios have been proposed to explain the critical 
behavior of the 3D XY spin-glass model.  As an attempt to clarify this controversy, we first study
the XYG model analyzing the spin correlation and integrated response functions, and 
the same chiral quantities.  Next, a similar analysis is performed for the XYB model. 

\subsection{XY spin-glass model with Gaussian bond distribution}

\begin{figure} [t!]
\includegraphics[width=6.5cm,clip=true]{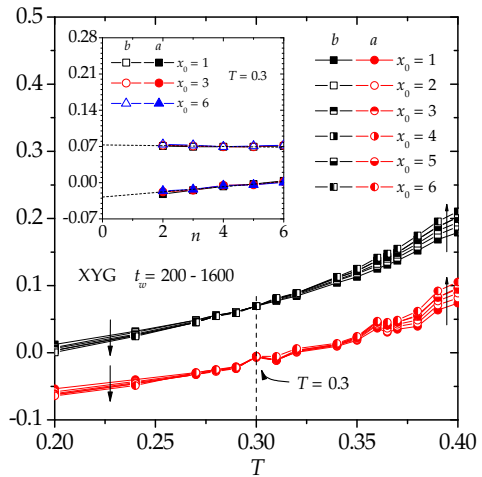}
\caption{\label{figure5} (Color online) The curves of the exponents $b$ and $a$ vs. $T$,
obtained for the XYG model.  As indicated, a range of $t_w = 200 - 1600$ and different values of $x_0$
were used.  Straight arrow indicate how the curves change with increasing $x_0$.
The inset shows the dependence of the exponents $b$ and $a$ whit $n$, 
for different values of $x_0$ at $T=0.3$.}
\end{figure}

For the XYG model we have studied a temperature range between $T=0.2$ and $0.4$.
Figure \ref{figure5} shows several sets of curves of the exponents $b$ and $a$ corresponding
to the spin degrees of freedom as a function of  $T$.
The curves shown were obtained fitting with different values of $x_0$ and with the waiting times 
$t_w = 200 - 1600$ ($n=4$). We clearly see in the plot that $b(T)\not=a(T)$, implying the
absence of a critical point for the spin degrees of freedom.
Data fitted using different ranges of $x$ (by changing $x_0$) show
an important dispersion in the values obtained for $b(T)$ and $a(T)$, as is displayed 
in Fig. \ref{figure5}. Such a dispersion with $x_0$ means that the assumed power law 
fit of Eqs.(\ref{scalingCorr}) and  (\ref{scalingResp}) is not actually followed in these cases.

However, we see that at $T=0.3$ the dispersion of the pseudo-exponents with $x_0$ is minimal
and both the curves $b(T,x_0)$ and the curves $a(T,x_0)$ have a crossing point in terms
of $x_0$ at this temperature. This effect is more clear  for the correlation exponent $b$,  
implying that a power-law scaling relation (\ref{scalingCorr})
is nearly followed in this case. In the inset of Fig. \ref{figure5}, we analyze in
detail the dependence of the pseudo-exponents with $x_0$ and with  different ranges of $t_w$ (parameter $n$)
at $T=0.3$. We see that $b$ is nearly independent of $x_0$ and weakly dependent on $n$. 
Extrapolating to the smallest $n$, we determine that it tends to $b=0.073(3)$. In the case of the 
pseudo-exponent $a$ we see that it has more fluctuations with $x_0$ and that it depends more strongly
on $n$, extrapolating to a negative value of $a=-0.028(5)$.    

\begin{figure} [t!]
\includegraphics[width=6.5cm,clip=true]{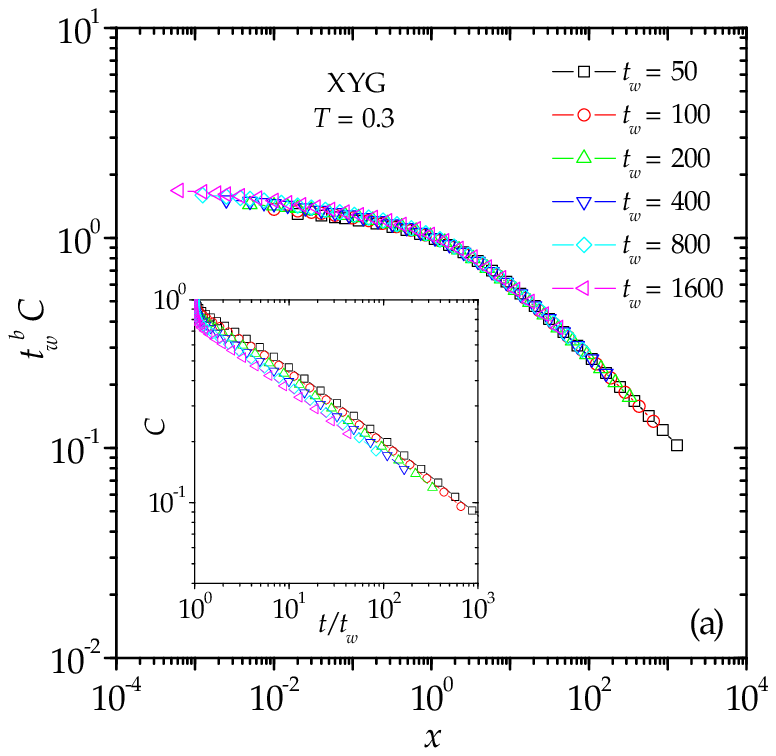}
\includegraphics[width=6.5cm,clip=true]{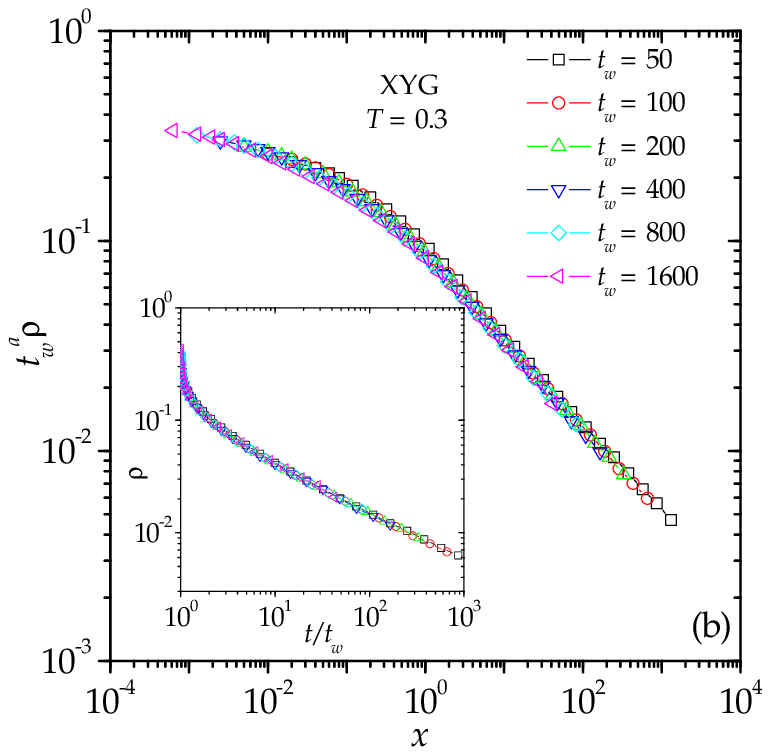}
\caption{\label{figure6} (Color online) XYG model.  Panels (a) and (b) show, respectively, 
the data collapse of the spin correlation and the spin integrated response functions at $T=0.3$
for different $t_w$ as indicated.  Collapses were obtained using $b=0.073$ and $a=-0.028$.   
Both insets show the corresponding curves as function of $t/t_w$.}
\end{figure}

In agreement with this previous analysis, Figs. \ref{figure6} (a) and (b) show good data collapses 
of the correlation and the integrated response functions for, respectively, $b=0.073$ and $a=-0.028$.  
However, we need to use a {\it negative} value for $a$ to obtain the
scaling of Eq.(\ref{scalingResp}).  If all waiting times are considered ($n=6$),
the inset in Fig. \ref{figure5} shows that $a \approx 0$.  Note in the inset of 
Fig. \ref{figure6} (b) that the curves of the response function do not depend appreciably on $t_w$. 
Even when the behavior of $b$ at this temperature implies a power law dependence
of the two-times correlation function as given by Eq.(\ref{scalingCorr}), the fact that clearly $b\not=a$ means
that, according to our method, this temperature can not be regarded as a critical one.

\begin{figure} [t]
\includegraphics[width=6.5cm,clip=true]{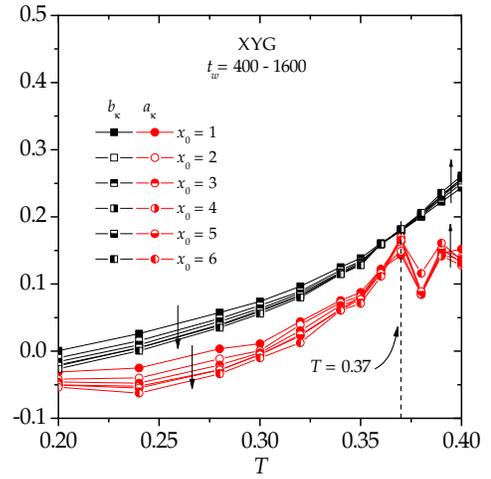}
\caption{\label{figure7} (Color online) The curves of the exponents $b_\kappa$ and $a_\kappa$ vs. $T$,
obtained for the XYG model.  As indicated, a range of $t_w = 400 - 1600$ and different values of $x_0$
were used.  Straight arrow indicate how the curves change with increasing $x_0$.}
\end{figure}

\begin{figure} [b]
\includegraphics[width=6.5cm,clip=true]{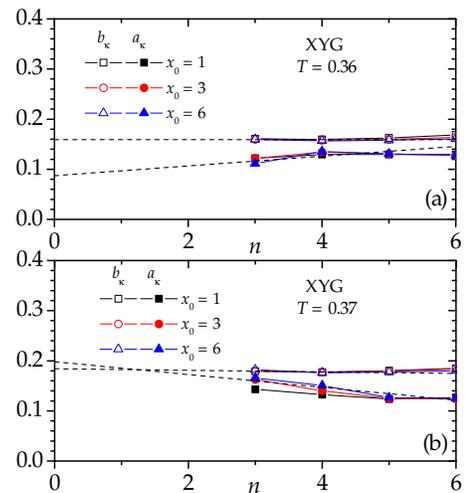}
\caption{\label{figure8} (Color online) XYG model.  The dependence of the exponents 
$b_\kappa$ and $a_\kappa$ whit $n$ for different values of $x_0$ as indicated, at
(a) $T=0.36$ and (b) $T=0.37$.}
\end{figure}

\begin{figure} [t!]
\includegraphics[width=6.5cm,clip=true]{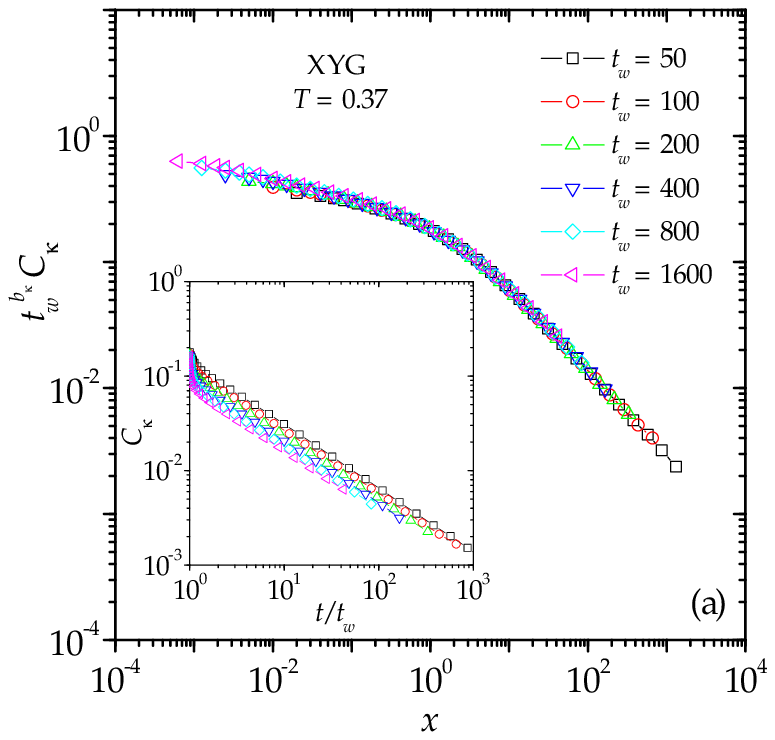}
\includegraphics[width=6.5cm,clip=true]{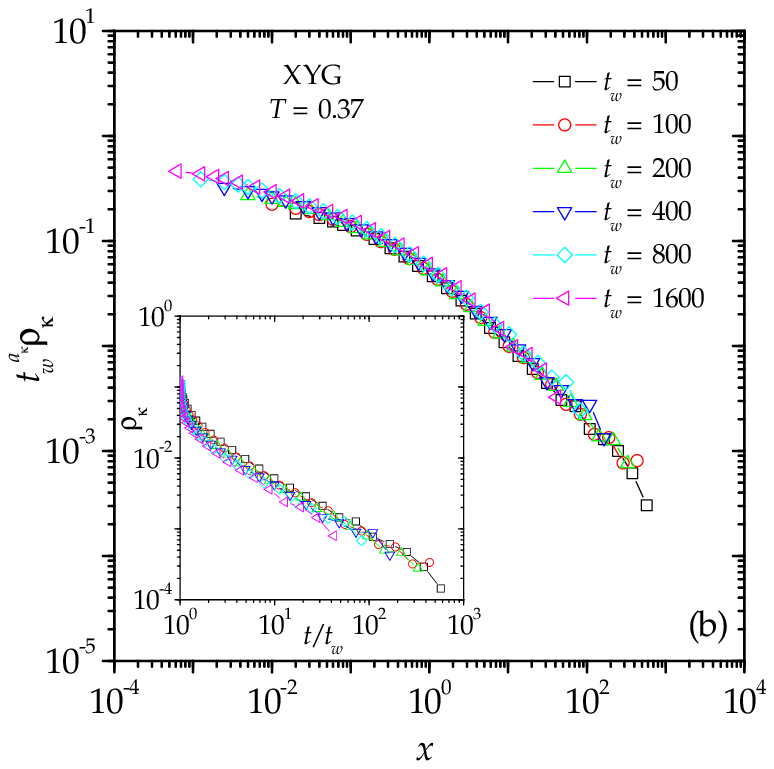}
\includegraphics[width=6.5cm,clip=true]{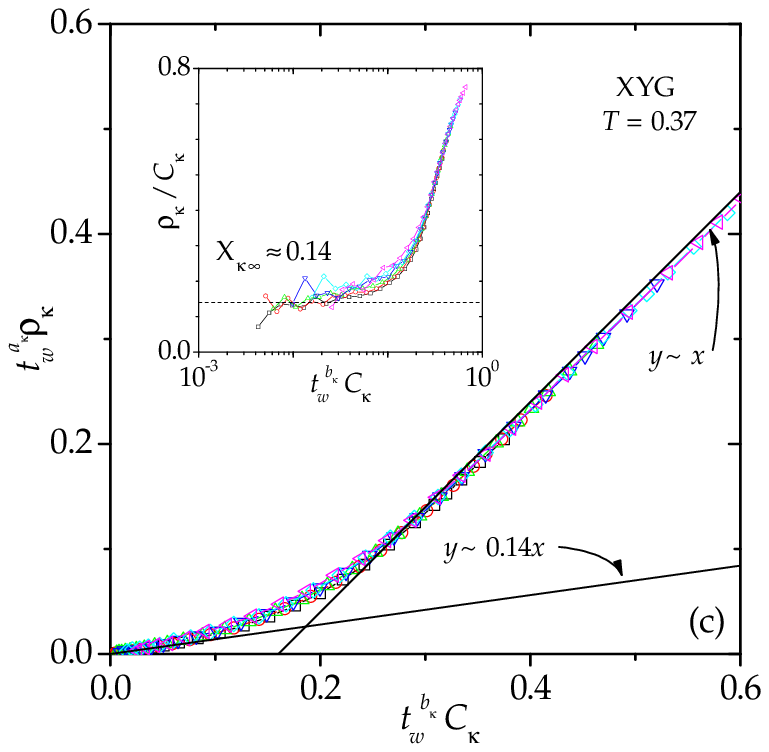}
\caption{\label{figure9} (Color online) XYG model. Panels (a) and (b) show, 
respectively, the data collapse with $b_\kappa=a_\kappa=0.19$ of the chiral correlation 
and the chiral integrated response at $T=0.37$ for different $t_w$ as indicated. 
Both insets show the corresponding curves as function of $t/t_w$.  (c) FDT plot.  
Inset shows the plot of $\rho/C$ vs. $t_w^b C$ and the estimated value of $X_\infty \approx 0.14$.}
\end{figure}

\begin{figure} [t!]
\includegraphics[width=6.5cm,clip=true]{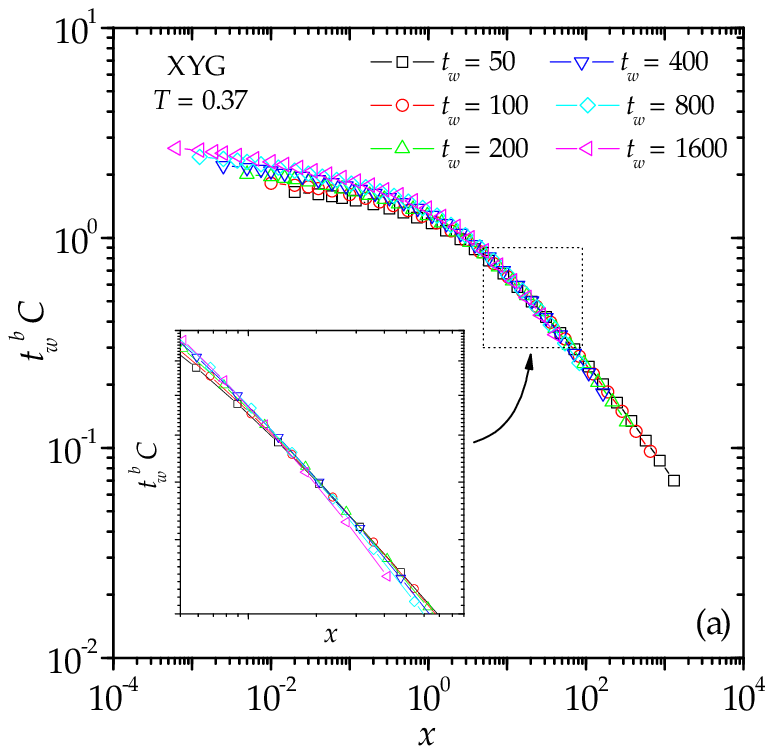}
\includegraphics[width=6.5cm,clip=true]{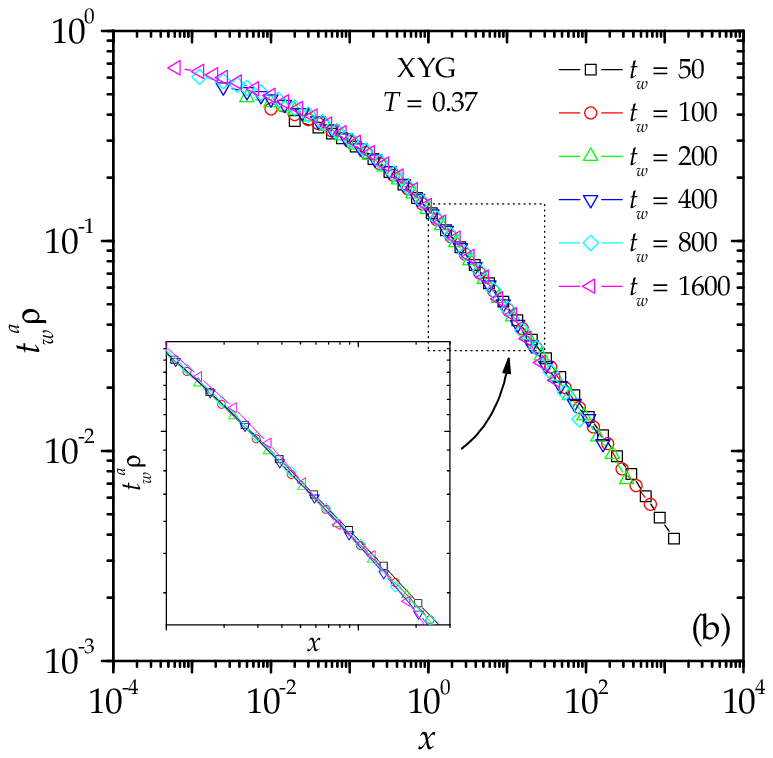}
\caption{\label{figure10} (Color online) XYG model.  Panels (a) and (b) show, respectively, 
the data collapse of the spin correlation and the spin integrated response at $T=0.37$
for different $t_w$ as indicated.  Collapses were obtained using $b=0.137$ and $a=0.035$.   
The insets show zooms of both set of curves.}
\end{figure}

We now turn to analyze the chiral quantities for the XYG model.   Figure \ref{figure7}
shows several sets of curves of $b_\kappa$ and $a_\kappa$ vs. $T$, which were obtained 
by choosing different values of $x_0$ and the ranges $t_w = 400-1600$ ($n=3$). 
In this case we see that there is a contact point of the two sets of curves
around at $T=0.37$ where $b_\kappa\approx a_\kappa$
showing the existence of a possible critical point.  However, we
note that in this case the curves of $a_\kappa$ show important fluctuations
with $x_0$ at $T=0.37$.  The situation becomes more clear            
when we consider the exponent values for different sets of $t_w$, as shown in    
Figs. \ref{figure8} (a) and (b) for, respectively, $T=0.36$ and $0.37$.  
Extrapolating, we see that for long waiting times, the condition $b_\kappa=a_\kappa$
can be fulfilled at a temperature close to $T=0.37$. This implies
that a phase transition involving the chiral degrees of freedom occurs at $T_c=0.37(1)$,
with  a common nonequilibrium exponent  $c_\kappa=0.19(1)$.  
To further confirm this finding, in
Figs. \ref{figure9} (a) and (b) we show the curves of the chiral correlation and
the chiral integrated response at $T=0.37$, where a value of $b_\kappa=a_\kappa=0.19$ was used 
to collapse simultaneously both sets of curves.  In addition, Fig. \ref{figure9} (c) shows 
the corresponding FDT plot and how for long times the FDT is violated with $X_\infty \approx 0.14$ . 
Note that this is not the best way to collapse the whole data set: 
in particular, in Fig. \ref{figure9} (b), we should have used $a_\kappa=0.12 $, which is the 
corresponding exponent value for $n=6$ [see Fig. \ref{figure8} (b)].
Instead, we have used a common value obtained extrapolating to smallest $n$. 
This allows to accomplish a collapsed FDT plot. If data for longer waiting times
were possible to obtain, we expect that the collapse in Fig. \ref{figure9} (b) could be
improved.

On the other hand, the correlation and integrated response functions of the spin degrees
of freedom have large fluctuations in both pseudo-exponent curves around $T=0.37$ in Fig.~\ref{figure5}.
Figures \ref{figure10} (a) and (b) show an attempt to collapse the correlation and 
the integrated response functions at this temperature using, respectively, $b=0.137$ and $a=0.035$
(two different values obtained for $n=4$ and $x_0=1$).  Analyzing the insets, which show 
zooms of both figures, we observe that the curves with different $t_w$ cross and then it is not possible 
to achieve a better data collapse by choosing another set of exponent values.  In fact, according to
Fig.~\ref{figure5}, the best data collapses for different values of $x_0$,
result in very different pairs of values of $b$ and $a$.  
Conversely, at $T=0.3$, where the $a$ and $b$ pseudo-exponents of the spin degrees of freedom
have minimum fluctuations, we see in Fig. \ref{figure7} that
$a_\kappa$ and $b_\kappa$ have fluctuations in their value when varying $x_0$.
Also, for this temperature, abnormal collapses (where both sets of curves 
cross for any pair of exponent values) are obtained for both the chiral
correlation and the chiral integrated response functions.    
All these analyses show that the spin and chiral degrees of freedom behave differently
in the XYG, at least in what regards to their nonequilibrium dynamics.

\subsection{XY spin-glass model with bimodal bond distribution}

\begin{figure} [t!]
\includegraphics[width=6.5cm,clip=true]{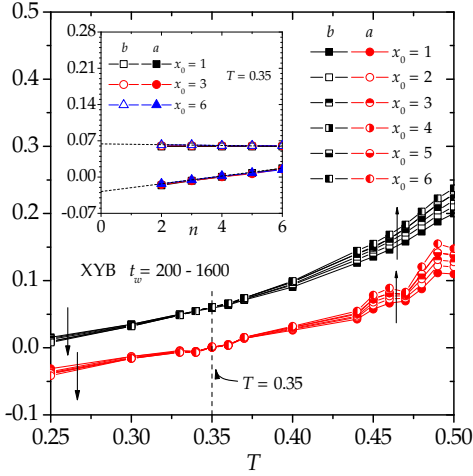}
\caption{\label{figure11} (Color online) The same plot as Fig. \ref{figure5} but for the XYB model.}
\end{figure}

We carried out for the XYB model a similar analysis as before, but now
for temperatures between $T=0.25$ and $0.54$.  We begin analyzing the spin quantities.
Figure \ref{figure11} shows several sets of curves of $b$ and $a$ vs. $T$,
which were obtained by choosing different values of $x_0$ and the ranges $t_w = 200 - 1600$. 
Again, we do not observe a temperature where the condition $b=a$ is fulfilled, but
there is a  point for the curves of $b$ (and $a$) at $T=0.35$ where their fluctuations
with $x_0$ is minimal.  The inset in Fig. \ref{figure11} 
shows that, by extrapolating to the smallest $n$, two very different values of 
$b=0.064(2)$ and $a=-0.028(3)$  are obtained.

\begin{figure} [t!]
\includegraphics[width=6.5cm,clip=true]{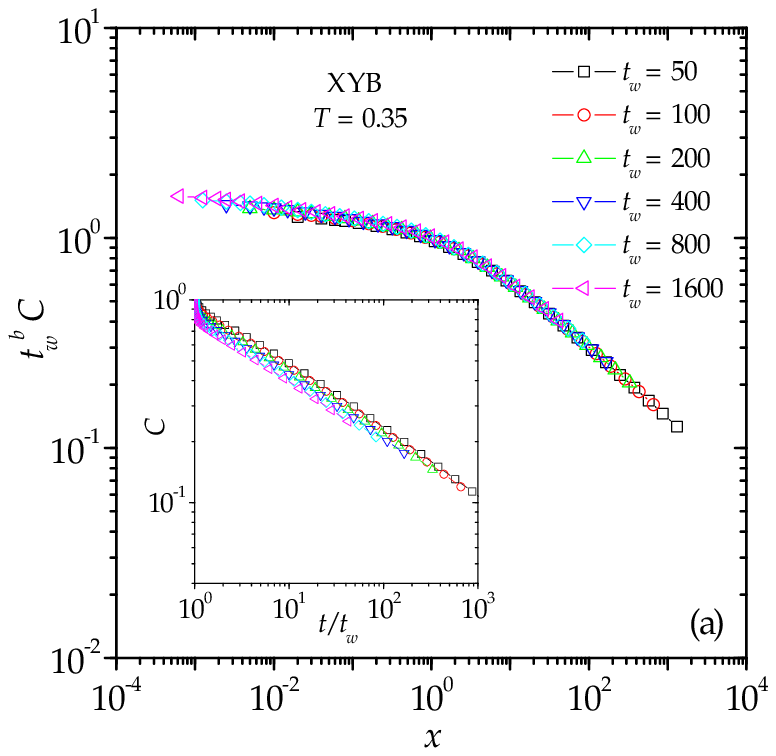}
\includegraphics[width=6.5cm,clip=true]{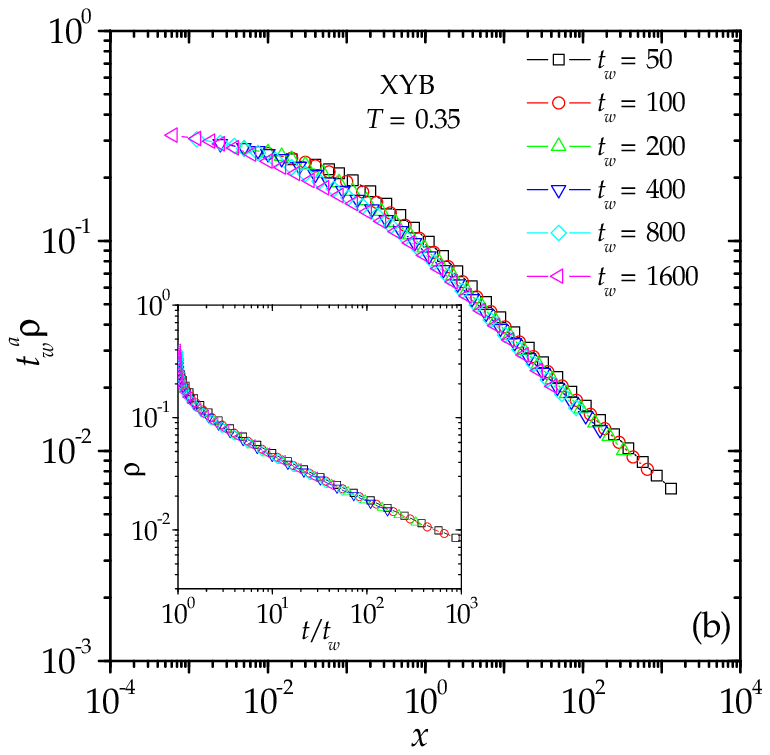}
\caption{\label{figure12} (Color online) The same plot as Fig. \ref{figure6} but for the XYB model 
at $T=0.35$ with $b=0.064$ and $a=-0.028$.}
\end{figure}

Since $b \ne a$, the temperature $T=0.35$ can not be regarded as corresponding to a
critical point. Figures \ref{figure12} (a) and (b) show the best data collapses 
of the correlation and the integrated response functions at $T=0.35$ obtained using, respectively, 
$b=0.064$ and $a=-0.028 <0$.   
The same behavior as found before for the XYG model at $T=0.3$ [see Figs. \ref{figure6} (a) and (b)].
Then, we also conclude that for the XYB model there is not a critical point for the spin degrees of freedom 
within the range of temperatures studied.   

\begin{figure} [t!]
\includegraphics[width=6.5cm,clip=true]{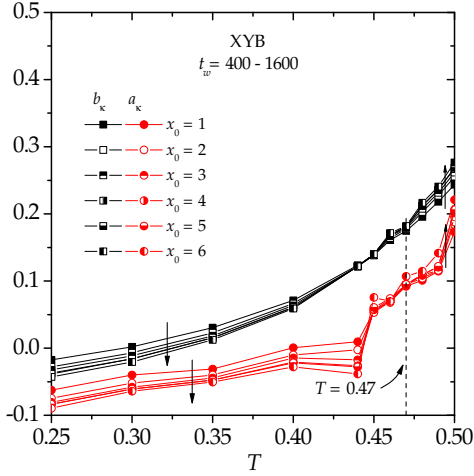}
\caption{\label{figure13} (Color online) The same plot as Fig. \ref{figure7} but for the XYB model.}
\end{figure} 

\begin{figure} [t!]
\includegraphics[width=6.5cm,clip=true]{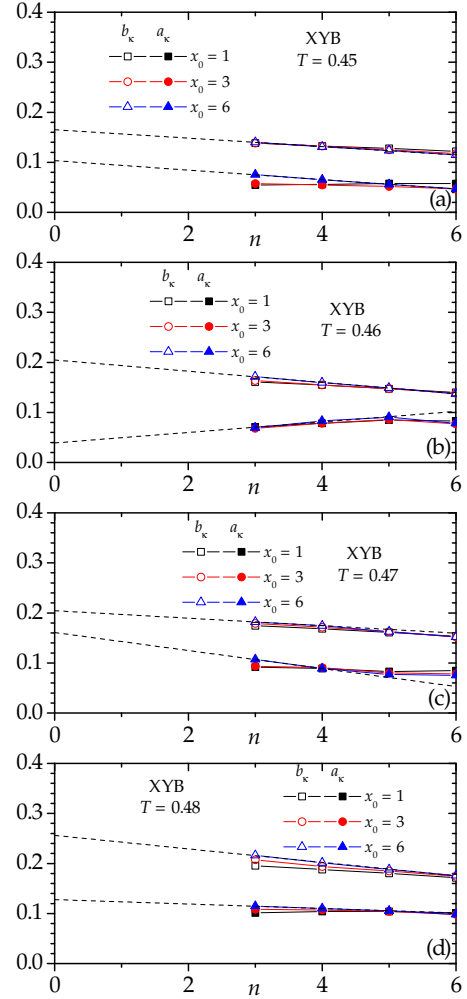}
\caption{\label{figure14} (Color online) The same plot as Fig. \ref{figure8} but for the XYB model
at (a) $T=0.45$, (b) $T=0.46$, (c) $T=0.47$ and (d) $T=0.48$.}
\end{figure}

\begin{figure} [t!]
\includegraphics[width=6.5cm,clip=true]{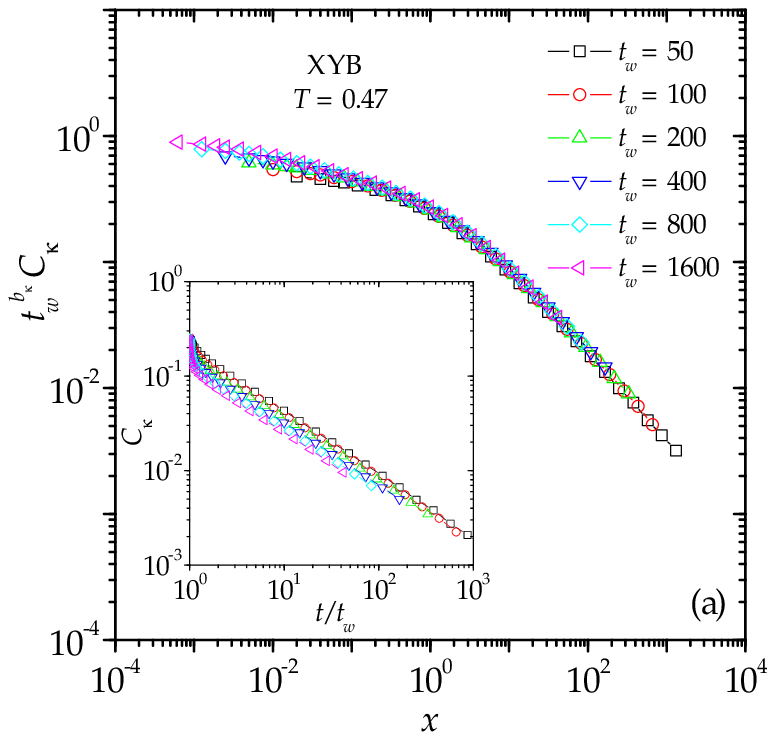}
\includegraphics[width=6.5cm,clip=true]{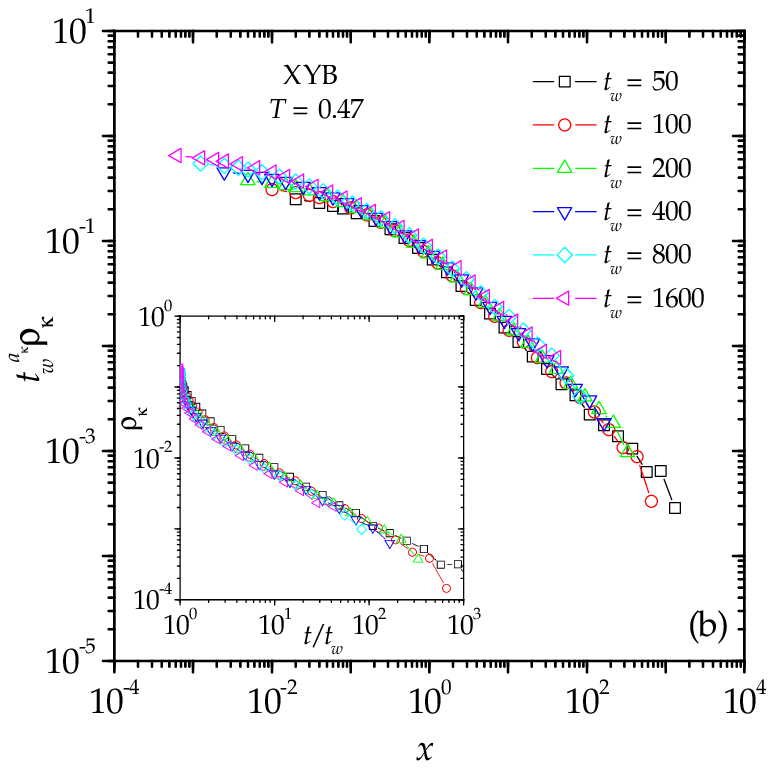}
\includegraphics[width=6.5cm,clip=true]{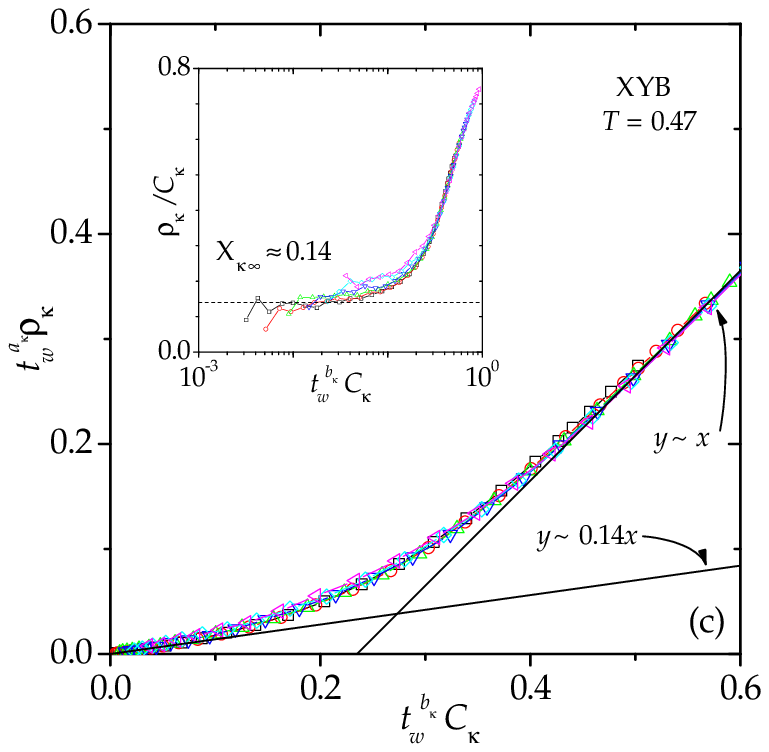}
\caption{\label{figure15} (Color online) The same plots as Figs. \ref{figure9} but for the XYB model
at $T=0.47$ and $b_\kappa=a_\kappa=0.18$.}
\end{figure}

Next, we present the results obtained for the chiral quantities.  Figure \ref{figure13} 
shows several sets of curves of $b_\kappa$ and $a_\kappa$ vs. $T$, which were obtained 
by choosing different values of $x_0$ and the ranges $t_w = 400 - 1600$ ($n=3$). 
We do not see, as in the previous case of the XYG, a point where $b_\kappa \approx a_\kappa$.
However, in the range $0.45<T<0.48$ the curves of $b_\kappa$ have small fluctuations with $x_0$ and
curves with different $x_0$ intersect at a temperature within this range.
The corresponding curves
for $a_\kappa$ also intersect for different $x_0$ in this region but with important fluctuations.  
We analyze in detail the behavior of the obtained pseudo-exponent values 
for different sets of $t_w$, as shown in Figs. \ref{figure14} (a)-(d), for temperatures between $T=0.45$ and 
$0.48$. From these data, we conclude that the calculation of the pseudo-exponents
needs larger $t_w$ to fully converge in this range of temperatures.
Comparing the behavior at different $T$,
it is possible to infer that the condition $b_\kappa=a_\kappa$ could be fulfilled at a 
temperature close to $T=0.47$. Extrapolating for the smallest $n$, 
we estimate that possibly there is a critical point for the chiral degrees of freedom 
at $T_c=0.47(1)$ with $c_\kappa=0.18(2)$. Even when the evidence for the existence of
a critical point is very weak as compared with the same case in the XYG, we
note that the obtained chiral nonequilibrium exponents $c_\kappa$ 
agree within the error bars for both spin-glass models.   
Figures \ref{figure15} (a) and (b) 
show the curves of the chiral correlation and the chiral integrated response at $T=0.47$, 
where a value of $b_\kappa=a_\kappa=0.18$ was used to collapse simultaneously both sets of curves.  
Figure \ref{figure15} (c) shows the corresponding FDT plot and the violation of FDT with 
$X_\infty \approx 0.14$.  

On the other hand, in Fig. \ref{figure11} we observe at $T=0.47$ large fluctuations in the pseudo-exponent curves
for the spin degrees of freedom. For this temperature the spin correlation 
and the spin response functions present abnormal data collapses similar to that shown previously for 
the XYG model in Figs. \ref{figure10} (a) and (b).  In addition, at
the temperature $T=0.35$, where the pseudo-exponents $b(T)$ and $a(T)$ have minimum fluctuations,
the corresponding curves $b_\kappa(T)$ and $a_\kappa(T)$ are very dependent on $x_0$ and
also abnormal data collapses  are obtained when attempted in this case.        
As before for the XYG model, we conclude that spin and chiral degrees of freedom
show different behavior in their nonequilibrium dynamics.

\section{\label{DC} DISCUSSION AND CONCLUSIONS}

In this work we have studied the equilibrium critical behavior of different 
three-dimensional XY models, by using a standard Monte Carlo scheme and a simple 
nonequilibrium method.\cite{Roma2010}  This technique relies on the calculation of the correlation 
and the integrated response functions at different temperatures and waiting times.  
Performing a data collapse analysis based on the scaling relations (\ref{scalingCorr}) 
and (\ref{scalingResp}), it is possible to determine the critical temperature and the
nonequilibrium scaling exponents.

First, we studied the well-known 3D ferromagnetic XY model to validate 
the nonequilibrium method when models with XY-type spins are simulated. The
results show that the numerical technique works well in such nondisordered system but,
as expected, this is not competitive when compared to the standard equilibrium 
and nonequilibrium methods to obtain $T_c$. In spite of this, we are
able to obtain a good estimate of the critical dynamical exponent $z_c$,
that improves previous reported calculations. We also studied the disordered 3D gauge-glass model.  
As discussed above, we determined a critical temperature of  
$T_c = 0.450(3)$ which is in good agreement with the values reported in the literature 
but more precise.  The nonequilibrium exponent $c = 0.070(4)$, however, 
is a little larger than the value of $b = 0.06(1)$ reported by us in a previous 
work \cite{Roma2008} using a more imprecise method (see above).

We have performed a more comprehensive study of the 3D XYG and XYB spin-glass models.
We find that the spin and chiral degrees of freedom behave very differently
in their relaxation dynamics in a wide range of temperatures.
When comparing their corresponding correlation and response functions
we find that spin and chiralities show different behaviors at
different temperatures. Regarding the spin degrees of freedom,
for the two spin-glass models (XYG and XYB) and in a wide 
range of temperatures, we find no evidence for the existence of a critical point.
Nevertheless at $T=0.3$ for the XYG model and at $T=0.35$ for the XYB model, where $b \ne a$, 
the spin correlation and the spin integrated response functions show good data collapses
to a power-law dependence (but, we remark, the corresponding chiral quantities do
not show power-law dependences at that $T$).
Regarding the chiral degrees of freedom,
for the XYG model we determined that there is a critical point at $T_c=0.37(1)$
with the common nonequilibrium exponent $c_\kappa=0.19(1)$.
However, in the case of the XYB model, the evidence for a chiral critical point is very weak,
and curves with longer waiting times are
necessary to clarify this issue (as it is evident from Fig. \ref{figure13}). 
We infer from our data the possibility of a critical point at $T_c=0.47(1)$ 
with $c_\kappa=0.18(2)$. 
Nevertheless, assuming that this estimated $T_c$ is correct, 
we should note that the exponent values of $c_\kappa$ agree within the error bars in
both models, suggesting that 
the universality class does not depend on the bond distribution form. 

As we discussed in the Introduction, some simulation studies found evidence in 
favor of a single finite critical temperature at which both symmetries simultaneously are broken. 
Specifically for the 3D XYG model, from equilibrium simulations and by analyzing the spin and the chiral correlations 
lengths, Lee and Young \cite{Lee2003} found that this occurs at $T_c=0.34(2)$ 
with the correlation exponent $\nu=1.2(2)$.  Also by studying the dynamical 
behavior of resistivity, Granato \cite{Granato2004a} reports $T_c=0.335(15)$, $z_c=4.5(3)$ and
$\nu=1.2(2)$, but found that the data are consistent with a phase transition in the range
$T_c \sim 0.25 - 0.35$.  More recently, Chen \cite{Chen2009} has obtained similar parameters,
$T_c=0.33(2)$, $z_c=4.0(1)$ and $\nu=1.4(1)$, by large-scale simulations performed 
with a resistive shunted-junction dynamics.  On the other hand, for the 3D XYB model 
Granato \cite{Granato2004b} reports $T_c=0.39(2)$, $z_c=4.4(3)$ and $\nu=1.2(2)$, but again
found that the data is consistent with a phase transition in the range $T_c \sim 0.3 - 0.45$.   
In this context, we remark that our findings for both spin-glass models indicate higher values of
the critical temperatures (beyond, of course, of the different scenario found by us).   

Within this same single-transition picture, Yamamoto {\em et al.} \cite{Yamamoto2004} using a 
nonequilibrium method, report for the 3D XYB model that both transitions occur simultaneously 
within a short range of temperatures, $T_c \sim 0.45 - 0.47$, and that the critical dynamic exponent is 
$z_c \approx 6$ and the exponent $\eta$ is close to zero. In this case the critical temperature
is very close to the value obtained by us. Meanwhile, Pixley and Young \cite{Pixley2008}
have proposed a more complex scenario in which the lower critical dimension could be close 
or equal to three, and then the finite-temperature phase transition should be removed by 
fluctuations.  They report, however, that for the 3D XYG model, the crossing temperature for the spin and the chiral 
correlations lengths moves towards $T \simeq 0.3$ when lattices of size up to $L=24$ are considered.     
Note that near this temperature, we found good data collapses of power-law 
scalings for dynamical correlation and response functions, 
but since $b \ne a$ in this case, we do not associate this temperature to a critical point.    

The possibility of a spin-chiral decoupling scenario has been suggested in the literature
mainly in the framework of equilibrium simulations.  In chronological order, Kawamura and Li \cite{Kawamura2001} 
found that the chiral critical temperature for the 3D XYB model is $T_c \approx 0.39$, $z_c=7.40(10)$, $\nu=1.2(2)$,
and $\eta=0.15(20)$.  On the other hand, very recently, Obuchi and Kawamura \cite{Obuchi2013} carried
out large-scale Monte Carlo  simulations of the 3D XYG model. There, they report two sets 
of critical parameters: for the spin-glass transition $T_c=0.275^{+0.013}_{-0.052}$,
$\nu=1.22^{+0.26}_{-0.06}$ and $\eta=-0.54^{+0.24}_{-0.52}$, and for the chiral-glass 
transition $T_c=0.313^{+0.013}_{-0.018}$, $\nu=1.36^{+0.15}_{-0.37}$ and $\eta=0.26^{+0.29}_{-0.26}$. 
Note that, although our findings agree in some sense with these works because we found strong evidence in 
favor of a decoupling scenario, the critical temperatures reported by us are very different. 

Another possible scenario is that there is no phase transition for
neither the spin nor the chiral degrees of freedom (as recently discussed
by Pixley and Young \cite{Pixley2008}) while, as found here, spins and
chiralities relax slowly with different dynamical behaviors at different temperatures.
This may explain the several discrepancies found in the literature both
regarding the coupling/decoupling issue as well as the value of the critical temperatures,
since differences in equilibration algorithms, system sizes and/or data analysis may lead
to different conclusions in such a case.
However, our results for the waiting times and system size ($L=50$) considered
support, at least for the XYG model, the existence of a chiral  
critical point at a finite temperature while the spins order at very low or zero temperature.

Finally, we remark the importance and simplicity of the method that we have used in this work. 
In a nonequilibrium simulation, many quantities (including the correlation and the integrated 
response functions) display power-law decays in a range of temperatures close to the critical one.
For disordered and frustrated systems, this range is very wide and then the standard methods fail to
determine correctly the value of $T_c$.\cite{Roma2010}  Here, the sensitivity of the nonequilibrium 
aging method is improved by requiring $b=a$, a condition which is expected to be fulfilled in a normal 
continuous-phase transition.\cite{Janssen1989,Calabrese2005}  This makes the method useful
for studying a broad range of systems with very slow dynamics. 

\begin{acknowledgments}
We thank L. F. Cugliandolo for fruitful discussions.
F. Rom\'a acknowledges financial support from CONICET (PIP 114-201001-00172), and Universidad Nacional de San Luis (PROIPRO 31712). D. Dom\'{\i}nguez acknowledges support from CNEA, CONICET (PIP11220090100051), and ANPCYT (PICT2011-1537).

\end{acknowledgments}
\appendix*
\section{}

Following the Refs.~\onlinecite{Chatelain2003,Ricci-Tersenghi2003,Berthier2007}, 
here we show how to calculate the integrated response to an infinitesimal external 
field.  

For a lattice site $i$, let us consider an arbitrary local scalar observable 
$\mathcal{A}_i$.  In a Monte Carlo dynamics (discrete-time Markov chain) the mean 
value of this quantity at time $t$ is given by
\begin{equation}\label{Smedio_0}
\langle \mathcal{A}_i(t) \rangle_0 = \sum_k \mathcal{A}_i^k(t) Q_0^k \left(t' \rightarrow t \right) P\left( C_{t'}^k , t' \right),
\end{equation}        
where the sum runs over all possible trajectories $k$ which occur between times 
$t'$ and $t$, $Q_0^k \left(t' \rightarrow t \right)$ is the probability of these 
trajectories in the unperturbed dynamics, $C_{t'}^k$ is the configuration of 
the system at time $t'$ for each trajectory, $\mathcal{A}_i^k(t)$ is the value 
of the observable at time $t$ in trajectory $k$ and $P\left( C_{t'}^k , t' \right)$ 
is the probability of that at time $t'$ the system is in the configuration $C_{t'}^k$ 
(note that here the capital letter $C$ does not represent the correlation function, 
but the configuration of the system).  If at time $s$ a trial configuration $q_s^k$ 
is attempted ($C_{s}^k \rightarrow q_s^k$) with acceptance rate $\omega_s^k$, then 
for the trajectory $k$ the transition probability in the unperturbed system, 
$W_0 \left( C_{s}^k \rightarrow C_{s+1}^k \right)$, is
\begin{equation}
W_0 \left( C_{s}^k \rightarrow C_{s+1}^k \right)= \delta_{C_{s+1}^k,q_s^k} \ \omega_s^k + \delta_{C_{s+1}^k,C_s^k}
\left( 1-\omega_s^k \right), \label{trans0}
\end{equation}
where $\delta$ is the delta Kronecker function.  In terms of (\ref{trans0}) 
the probability $Q_0^k$ can be written as
\begin{equation}
Q_0^k \left( t'\rightarrow t \right) = \prod_{s=t'}^{t-1} W_0
\left( C_{s}^k \rightarrow C_{s+1}^k \right).
\end{equation}

On the other hand, if a local perturbation (coupled to the observable $\mathcal{A}_i$) 
of strength $h_i$ is on between times $t_1$ and $t_2$, then the mean value of 
$\mathcal{A}_i$ will be
\begin{eqnarray}
\langle \mathcal{A}_i(t) \rangle_h =  \sum_k && \mathcal{A}_i^k (t) Q_0^k \left(t' \rightarrow t_1 \right) Q_h^k \left( t_1 \rightarrow t_2
\right) \nonumber \\ 
 && \times Q_0^k \left( t_2 \rightarrow t \right) P\left( C_{t'}^k ,t' \right), \label{mean_value_h}
\end{eqnarray}
where now
\begin{equation}
Q_h^k \left( t_1 \rightarrow t_2 \right) = \prod_{s=t_1}^{t_2-1} W_h \left( C_{s}^k \rightarrow C_{s+1}^k \right)  ,
\end{equation}
and $W_h \left( C_{s}^k \rightarrow C_{s+1}^k \right)$ is the probability of 
transition in the perturbed simulation.     
To calculate the integrated response, we need to derive the mean value in 
(\ref{mean_value_h}) with respect to $h_i$     
\begin{eqnarray} 
\frac{ \partial \langle \mathcal{A}_i(t) \rangle_h}{\partial h_i} = \sum_k && \mathcal{A}_i^k (t) Q_0^k \left( t' \rightarrow t_1 \right)
\frac{ \partial Q_h^k \left( t_1 \rightarrow t_2 \right)}{\partial h_i} \nonumber \\ 
&& \times Q_0^k \left( t_2 \rightarrow t \right) P\left( C_{t'}^k , t' \right), \label{der1}
\end{eqnarray}
where
\begin{equation}
\frac{ \partial Q_h^k \left( t_1 \rightarrow t_2 \right)}{\partial h_i} = Q_h^k \left( t_1 \rightarrow t_2 \right) \sum_{s=t_1}^{t_2-1}  \frac{ \partial \ln \left[ W_h \left(C_{s}^k \rightarrow C_{s+1}^k \right) \right] }{\partial h_i}.
\end{equation}
In the limit  $h_i \to 0$ we define the function
\begin{equation} 
R_i^k \left( t_1 \rightarrow t_2 \right) = \sum_{s=t_1}^{t_2-1} \frac{ \partial \ln \left[ W_h \left( C_{s}^k \rightarrow C_{s+1}^k \right) \right] }{\partial h_i} \Bigg|_{h=0} , \label{der2}
\end{equation}
and therefore (\ref{der1}) can be written as
\begin{eqnarray} \label{der3}
\frac{ \partial \langle \mathcal{A}_i(t) \rangle_h}{\partial h_i} \Bigg|_{h=0} & =&  \sum_k \mathcal{A}_i^k (t) R_i^k \left( t_1 \rightarrow t_2 \right) Q_0^k \left( t' \rightarrow t \right) P\left( C_{t'}^k , t' \right) \nonumber \\
&=& \langle \mathcal{A}_i (t) R_i \left( t_1 \rightarrow t_2 \right) \rangle_0.
\end{eqnarray}
Particularly for a thermoremanent process, one chooses $t'= t_1 = 0$ and $t_2 = t_w$.  Finally, 
the reduced integrated response for a $N$-site system can be calculated by
\begin{eqnarray} 
\rho_\mathcal{A} (t,t_w) &=& \frac{T}{N} \sum_{i=1}^N \frac{ \partial \langle \mathcal{A}_i(t) \rangle_h}{\partial h_i} \Bigg|_{h=0} \nonumber \\
&=& \frac{T}{N} \sum_{i=1}^N \langle \mathcal{A}_i (t) R_i \left(0 \rightarrow t_w \right) \rangle_0 .\label{resp0}
\end{eqnarray}
Note that for a single spin dynamics, the derivative with respect to $h_i$ in 
(\ref{der2}) is nonzero only when, in the transition $C_{s}^k \rightarrow C_{s+1}^k$, the configurational change involve the site $i$.  It is important to stress that the mean values 
in (\ref{resp0}) are calculated in an unperturbed simulation.   

Thus, by adding to the Hamiltonian the perturbation term (\ref{perS}), it is 
easy to show that the reduced integrated response (\ref{respS}) can be 
calculated by means of the expression  
\begin{eqnarray}
\rho (t,t_w) = \frac{T}{N} \sum_{i=1}^N && \langle \cos \left( \theta_i (t) \right) R_{x,i} \left( 0 \rightarrow t_w \right) \rangle_0 \nonumber \\
  + && \langle\sin \left(\theta_i (t) \right) R_{y,i} \left( 0 \rightarrow t_w \right)\rangle_0 ,
\end{eqnarray}
where 
\begin{equation} 
R_{x/y,i}^k \left( 0 \rightarrow t_w \right) = \sum_{s=0}^{t_w-1} \frac{ \partial \ln \left[ W_h \left( C_{s}^k \rightarrow C_{s+1}^k \right) \right] }{\partial h_{x/y,i}} \Bigg|_{h=0} . \label{RXY}
\end{equation}
In the Monte Carlo simulation, when a local change $\Delta \theta_{i}$ is attempted, 
the initial and final phases are, respectively, $\phi_1 = \theta_i^k (s)$ and 
$\phi_2 = \theta_i^k (s) + \Delta \theta_i$. If the change is accepted with the 
Glauber rate (\ref{rate}), then the $s$-th terms in (\ref{RXY}) are
\begin{equation}
\frac{ \partial \ln \left[ W_h \right] }{\partial h_{x,i}} \Bigg|_{h=0} = \beta  \left[ \cos \left( \phi_2 \right) - \cos \left(\phi_1 \right)\right] (1-\omega)
\end{equation}
and
\begin{equation}
\frac{ \partial \ln \left[ W_h \right] }{\partial h_{y,i}} \Bigg|_{h=0}= \beta  \left[ \sin \left( \phi_2 \right) - \sin \left(\phi_1 \right)\right] (1-\omega) .
\end{equation}
On the other hand, if the change is rejected, $\theta_i^k (s+1) = \theta_i^k (s)$ 
and the corresponding terms are
\begin{equation}
\frac{ \partial \ln \left[ W_h \right] }{\partial h_{x,i}} \Bigg|_{h=0} = \beta  \left[ \cos \left( \phi_1 \right) - \cos \left(
\phi_2 \right)\right] \omega
\end{equation}
and
\begin{equation}
\frac{ \partial \ln \left[ W_h \right] }{\partial h_{y,i}} \Bigg|_{h=0} = \beta  \left[ \sin \left( \phi_1 \right) - \sin \left(
\phi_2 \right)\right] \omega .
\end{equation} 

In a similar way, by adding to the Hamiltonian the perturbation term (\ref{perC}), 
the reduced chiral integrated response (\ref{respC}) can be calculated by means of 
\begin{equation}
\rho_\kappa(t,t_w) = \frac{T}{3N} \sum_\alpha  \langle \kappa_\alpha(t) R_\alpha \left( 0 \rightarrow t_w \right) \rangle_0,  
\end{equation}
where 
\begin{equation} 
R_\alpha \left( 0 \rightarrow t_w \right) = \sum_{s=0}^{t_w-1} \frac{ \partial \ln \left[ W_f \left( C_{s}^k \rightarrow C_{s+1}^k \right) \right] }{\partial f_\alpha} \Bigg|_{f=0} . \label{RXY_C}
\end{equation}
Again, for a chirality change of $\Delta \kappa_\alpha$ the initial and final 
chiralities are, respectively, $\kappa_1 = \kappa_\alpha^k (s)$ and $\kappa_2 = \kappa_\alpha^k (s) + \Delta \kappa_\alpha$.  
If the change is accepted then the $s$-th term in (\ref{RXY_C}) is
\begin{equation}
\frac{ \partial \ln \left[ W_f \left( C_{s}^k \rightarrow C_{s+1}^k \right) \right] }{\partial f_\alpha} \Bigg|_{f=0} = \Delta \kappa_\alpha (1-\omega),
\end{equation}  
but if the change is rejected, $\kappa_\alpha^k (s+1) = \kappa_\alpha^k (s)$ 
and the corresponding term is
\begin{equation}
\frac{ \partial \ln \left[ W_f \left( C_{s}^k \rightarrow C_{s+1}^k \right) \right] }{\partial f_\alpha} \Bigg|_{f=0} = -\Delta \kappa_\alpha \omega.
\end{equation}  


\end{document}